\documentclass[12pt]{iopart}
\usepackage{iopams}

\begin{document}

\title[What is the $B$ mode]{What is the $B$ mode in CMB polarization?}

\author{Jack Y.~L.~Kwok$^1$, King Lau$^2$, Edward Young$^3$ and Kenneth
Young$^1$}

\address{$^1$ Department of Physics, The Chinese University of Hong Kong,
Hong Kong, China}
\address{$^2$ Department of Physics,
University of Minnesota, Minneapolis, MN 55455, USA}
\address{$^3$ Department of Physics, Stanford University, Stanford, CA 94305, USA}
\ead{jackkwok@link.cuhk.edu.hk, kennylau@umn.edu,
eyy@stanford.edu,  kyoung@cuhk.edu.hk}

\vspace{4mm}\noindent{\it Keywords}: CMB, polarization, B mode,
inflation

\vspace{4mm}

%
%
%

\begin{indented}
\item[]\today
\end{indented}


\begin{abstract}

A linear polarization field on a surface is expressed in terms of
scalar functions, providing an invariant separation into two
components; one of these is the $B$ mode, important as a signature
of primordial gravitational waves, which would lend support to the
inflation hypothesis.
The case of a plane already exhibits the key ideas, including the
formal analogy with a vector field decomposed into gradient plus
curl, with the $B$ mode like the latter.
The formalism is generalized to a spherical surface using
cartesian coordinates. Analysis of global data provides a path to
vector and tensor spherical harmonics.

\end{abstract}


%
%


\def\nid{\noindent}
\def\bsub{\begin{subequations}}
\def\esub{\end{subequations}}
\def\beq{\begin{equation} }
\def\eeq{\end{equation}}
\def\eeql#1{\label{#1} \end{equation}}
\def\bea{\begin{eqnarray}}
\def\eea{\end{eqnarray}}
\def\eeal#1{\label{#1} \end{eqnarray}}
\def\eean{\nonumber\end{eqnarray}}
\def\bed{\vspace{-4mm}\begin{displaymath}}
\def\eed{\end{displaymath}\vspace{-4mm}}
\def\bibspace{\hspace{2mm}}
\def\parvsep{\vspace{2mm}}

\def\om{\omega}
\def\Om{\Omega}
\def\ep{\epsilon}
\def\p{\partial}
\def\vxi{\mbox{\boldmath $\xi$}}
\def\vnabla{\mbox{\boldmath $\nabla$}}
\def\YLM{Y_{\ell m}}


\section{Introduction}
\label{sect:intro}

\subsection{Physical context}

The isotropic black-body Cosmic Microwave Background
(CMB)~\cite{penzias-1, penzias-2, dicke} indicates a dense and hot
early universe and launched modern big-bang cosmology. With a
dipole angular dependence~\cite{fixsen-1, fixsen-2} due to
observer motion ($v/c \sim 10^{-3}$) removed, the remaining
anisotropies are at the ${\sim} 10^{-4}$ level~\cite{cobe-first}
even across widely separated directions, thus prompting the
inflation hypothesis~\cite{guth, linde}. Better data~\cite{wmap}
have enabled precision cosmology.
Recent attention has turned to polarization~\cite{kogut},
conventionally separated into the so-called $E$ and $B$ modes, at
relative levels of ${\sim}10^{-6}$ and ${\sim}10^{-7}$. %
The CMB $E$ modes, with sharper acoustic peaks, provide additional
constraints on the standard Lambda cold dark matter ($\Lambda$CDM)
model parameters and help to determine the optical depth to
re-ionization.

The tiny $B$ mode is important in cosmology: to first order it
cannot be produced by scalar (e.g., density) perturbations, but
only by tensor perturbations~\cite{seljak, kamkos}, i.e., strong
gravitational waves. Thus a cosmological $B$ mode would support
the inflation hypothesis. However, detected signals so
far~\cite{bicep2-1} are not above foreground
contamination~\cite{bicep-planck, planck-foreground} (including
lensing from $E$ modes). Many further experiments are in
progress~\cite{progress} to pin down the cosmological $B$ mode.
Setting aside the enormous observational challenges (e.g.,
instrumental, low signal-to-noise, foreground contamination),
there is a key conceptual question that deserves to be better
understood by a wider community: How is the $B$ mode defined and
hence extracted from the polarization data?

The $E$ and $B$ modes are heuristically likened to a gradient and
a curl, or TE/ TM modes. Typical statements refer to ``a twisting
pattern, or spin, known as a curl or $B$-mode"~\cite{cowan}; or to
``the curl component ($B$ mode)"~\cite{kam1}; or to components
having respectively ``vanishing curl" and ``vanishing
divergence"~\cite{bunn}. If interpreted naively, these statements
are misleading --- the usual conception of gradient and curl does
not apply to polarization. This paper explains the decomposition
of the polarization field in an elementary way. By avoiding the
apparatus of curved manifolds (celestial sphere) and in particular
emphasizing the case of a plane, the account should be accessible
to a wide audience.

\subsection{Polarization}

Consider a wave propagating along $+z$, with electric field
components $E_j$, $j = 1, 2$, in the usual complex notation.
Quadratic averages $S_{ij} = \mbox{Av } (E^*_i E_j)$ form a $2
\times 2$ Hermitian matrix with $S_{11} = I+Q$, $S_{22} = I-Q$,
$S_{12} = S_{21}^* = U+iV$.
Of the four Stokes parameters~\cite{jack2}, the overall intensity
$I$ is considered separately. In CMB, $V$ is usually ignored,
because
primordial circular polarization is not expected in standard
models~\cite{soma, king};
with few exceptions that place upper limits~\cite{chuss, class},
$V$ is assumed to be zero as a systematic check~\cite{king}.
With $I, V$ removed, polarization is described by the symmetric
traceless tensor
\bea \mathbb{P} &=&
\left(%
\begin{array}{cc}
  Q  &  U \\
  U  & -Q  \\
\end{array}%
\right) \; . \eeal{eq:P-tens}

All the information in $\mathbb{P}$ is contained in its first
column ${\bf P} = (P_1 , P_2)^{\rm T} = ( Q , U )^{\rm T}$, to be
called a doublet. The linear polarization of magnitude $|{\bf P}|$
is along the direction $\chi$, with $\tan 2 \chi = U/Q$. Upon
rotation by an angle $\alpha$, $Q \pm iU$ gain a factor $e^{\pm
2i\alpha}$, i.e., they transform as spin $\pm 2$.

\subsection{The central issue}

Polarization $\mathbb{P}$, a rank-2 tensor, cannot be ``similar
to" a vector ${\bf V}$, for which one speaks of gradient and curl.
A vector field (arrows) returns to itself upon rotation by $2\pi$.
A linear polarization field (\textit{unsigned} line segments)
returns to itself upon rotation by $\pi$. There is no polarization
analogue to $\int {\bf V}\cdot d{\bf r}$.

If a scalar field varies along ${\bf k}$, then a vector field that
is a gradient (curl) makes an angle $n \pi/2$ with ${\bf k}$, with
$n$ even (odd). However, $E$ ($B$) mode polarization makes an
angle of $n \pi/4$ with ${\bf k}$, with $n$ even (odd).

Thus a factor of 2 appears in the rank of the tensor, the phase
under rotations and the angle relative to ${\bf k}$ --- all for
the same reason. To avoid the strained analogy, henceforth the two
components will be simply called the $A$ and $B$ modes and denoted
with superscripts ${}^a$ and ${}^b$.


The mathematical formalism in the literature~\cite{seljak, bunn}
that supports the heuristic analogy is typically bundled with
additional elements. First, reference to inflation~\cite{kam1} is
not needed, just as an electric field can be decomposed into a
gradient plus a curl without reference to Maxwell's equations. The
complications of a sphere~\cite{bunn} can be avoided by (first)
considering a plane, useful for narrow-field
observations~\cite{bicep1}
or as a stepping stone to a sphere. A global harmonic
analysis~\cite{bunn} is also not strictly necessary: the
separation can in principle be implemented locally. This paper
strips away these complications to provide a simple account of the
separation of a linear polarization field into two components, for
access by a wider audience.

\subsection{Overview}

The paper is structured so that readers can stop at various points
depending on the level of technical detail desired. The formalism
is first developed for the $xy$ plane using the polar
representation (Section~\ref{subsect:plane-pol}): a vector field
is expressed as $(\p_1 + i \p_2)$ and a polarization field as
$(\p_1 + i \p_2)^2$ acting on scalar fields, with equations
(\ref{eq:2-01}) and (\ref{eq:2-11}) exhibiting the analogy. As far
as concepts are concerned, this is the end of the story.

The formalism, rewritten in cartesian coordinates
(Sect.~\ref{subsect:plane-cart}), is ported first to a plane with
unit normal $\hat{\bf n}$ (Sect.~\ref{subsect:another}) and then,
upon $\hat{\bf n} \mapsto {\bf r}$, to a \textit{small} part of a
unit sphere (Sect.~\ref{sect:sphere}). The resultant formalism,
involving the angular momentum ${\bf L}$ rather than $\vnabla$,
applies ``classically", i.e., if commutators are ignored, valid
for mode number $\ell \gg 1$. The proper ordering of operators
yields lower-order corrections in ${\bf L}$. Global analysis
(Sect.~\ref{sect:globe}) leads to vector and tensor spherical
harmonics, and provides a tool for disentangling different
contributions to each polarization component.



\section{Formalism for a plane}
\label{sect:plane}

The objective is to express either a vector field ${\bf V}$ or a
polarization field $\mathbb{P}$ on a surface as derivatives of two
scalar fields.

\subsection{Polar representation}
\label{subsect:plane-pol}

\subsubsection{Vector case.}
The decomposition of a vector field into a gradient and a curl
(Helmholtz's theorem) need not be repeated, but there are special
features in two dimensions. Given a real vector field ${\bf V} =
(V_1 , V_2)^{\rm T}$ on the $xy$ plane (say in the domain $0 \le x
\le {\cal L}_x$, $0 \le y \le {\cal L}_y$ with periodic boundary
conditions), we seek a representation (the only form consistent
with the angular momentum structure)
\bea V_1 \pm i V_2 &=& (\partial_1 \pm i \partial_2) \, (\Phi^a
\pm i \Phi^b) \eeal{eq:2-01}
in terms of real scalar functions $\Phi^{a,b}$. Henceforth only
one of the conjugate equations will be shown.
To prove such a representation, consider
\bea \varphi^a + i \varphi^b &\equiv& (\partial_1 -i \partial_2 )
\, (V_1 + i V_2) \; . \eeal{eq:2-02}
Since $(\p_1 -i \p_2)$ has $-1$ unit of angular momentum,
$\varphi^{a,b}$ are scalars. The key identity
\bea (\partial_1 - i \partial_2 ) \, (\partial_1 + i \partial_2)
&=& \nabla_\perp^2 \eeal{eq:2-03}
leads to
\bea \varphi^a + i \varphi^b &=& \nabla_\perp^2 \, (\Phi^a + i
\Phi^b) \; . \eeal{eq:2-04}

Thus, given ${\bf V}$, first finds (locally) $\varphi^{a,b}$ by
(\ref{eq:2-02}); then invert the laplacian to give $\Phi^{a,b}$.
With global data, the inversion is unique, proved by expanding in
the eigenfunctions of $\nabla_\perp^2$ --- provided
$\varphi^{a,b}$ have no projection on the null space of
$\nabla_\perp^2$. The latter condition is obvious from
(\ref{eq:2-02}) and the periodic boundary condition; another proof
is given in (\ref{eq:formal}) below. If data are confined to a
subset, one can imagine any extension of $\varphi^{a,b}$ to the
entire plane, and then invert; different extensions yield
different $\Phi^{a,b}$.

\subsubsection{Tensor case.}
Since $Q \pm i U$ have $\pm 2$ units of angular momentum, the
analogous representation must take the form
\bea Q \pm i U &=& (\partial_1 \pm i \partial_2)^2 \, (\Phi^a \pm
i \Phi^b) \; , \eeal{eq:2-11}
involving a second-order operator. Define
\bea \varphi^a + i \varphi^b &\equiv& (\partial_1 -i \partial_2
)^2 \, (Q + i U) \; . \eeal{eq:2-12}
Using the square of the identity (\ref{eq:2-03}), we find
\bea \varphi^a + i \varphi^b = K \, (\Phi^a + i \Phi^b) \; ,
\eeal{eq:2-14}
where $K$ and its eigenvalue $N_K$ (for plane waves with wave
number ${\bf k}$) are
\bea K = \nabla_\perp^4 \quad , \quad N_K = k^4 \; .
\eeal{eq:2-22K}
The algorithm is the same: use (\ref{eq:2-12}) to obtain (locally)
$\varphi^{a,b}$ and invert $K$ to obtain $\Phi^{a,b}$.

Thus polarization on a plane is expressible in terms of two
scalars, which importantly do not depend on the coordinate
choice~\cite{seljak}, in a manner analogous to the case of vector
fields; as immediately shown in (\ref{eq:2-gc}), the latter is the
same as a gradient plus a curl.

\subsubsection{Local separation.}
Whether for the vector or the tensor case, one can determine
$\varphi^{a,b}$ \textit{locally} and in principle assess the
relative importance of the $B$ mode on any subset $\Om$ of the
plane by the ratio $R^b/R^a$, where
$R^{a,b} = \int_\Om dS \, |\varphi^{a,b}|^2$.
%

\subsection{Cartesian representation}
\label{subsect:plane-cart}


\subsubsection{Vector case.}
In cartesian coordinates, (\ref{eq:2-01}) is equivalent to
\bea {\bf V} &=& {\bf d} \Phi^a - \tilde{\bf d}\Phi^b  \label{eq:2-01a} \\
&=& \vnabla_\perp \Phi^a + \vnabla_\perp \times (-\hat{\bf
n}\Phi^b) \; , \label{eq:2-gc} \eea
i.e., as a gradient plus a curl, where
\bea {\bf d} = \vnabla_\perp \quad , \quad \ \tilde{\bf d} = {}-
\hat{\bf n} \times {\bf d} \; , \eeal{eq:tilde-def}
in which $\hat{\bf n} =  \hat{\bf e}_z$ is the unit normal to the
plane. Duals can equivalently be defined by
\bea \tilde{\bf d} = \mathbb{E} {\bf d} \quad , \quad
\mathbb{E} \equiv \left(%
\begin{array}{cc}
  0 & 1 \\
  -1 & 0 \\
\end{array}%
\right) \; . \eeal{eq:dual}
This latter definition for duals also applies to doublets and
tensors.

The key identity (\ref{eq:2-03}) is equivalent to
\bea {\bf d}^\dagger \cdot {\bf d} = \tilde{\bf d}^\dagger \cdot
\tilde{\bf d} = - \nabla_\perp^2 \quad , \quad \tilde{\bf
d}^\dagger \cdot {\bf d} = 0 \; . \eeal{eq:2-13}
The orthogonality relation allows different components to be
projected out.
The definition of $\varphi^{a,b}$ in (\ref{eq:2-02}) becomes
\bea \varphi^a = -{\bf d}^\dagger \cdot {\bf V} = \vnabla_\perp
\cdot {\bf V}\quad , \quad \varphi^b = \tilde{\bf d}^\dagger \cdot
{\bf V} = \left( \vnabla_\perp \times {\bf V} \right) \cdot
\hat{\bf n} \; . \eeal{eqL2-14}

It can now be proved more formally that $\varphi^{a,b}$ have no
projection on the null space ${\cal S}$ of $\nabla_\perp^2$.
Consider $F \in {\cal S}$ and define inner products in the usual
way; then, e.g.,
\bea (F , \varphi^a) = (F , \vnabla_\perp \cdot {\bf V} ) &=& -
(\vnabla_\perp F , {\bf V} )  \; . \eeal{eq:formal}
But
$( \vnabla_\perp F , \vnabla_\perp F ) = (F , -\nabla_\perp^2 F )
= 0 $.
This proof generalizes to tensors, and to the case of a sphere
(for which $\varphi^{a,b}$ must have no projection on $\ell = 0,
1$).

\subsubsection{Tensor case.}
Write the second-order differential operator as
\bea (\partial_1 + i \partial_2)^2 &= &(\partial_1^2 -
\partial_2^2) + i (2 \partial_1\partial_2) \equiv D_1 + i D_2 \; . \nonumber \eea
%
Define
\bea \mathbb{D} &=& \left(%
\begin{array}{cc}
  D_1 & D_2 \\
  D_2 & -D_1 \\
\end{array}%
\right) \eeal{eq:2-17}
and its dual $\tilde{\mathbb{D}} = \mathbb{E}\mathbb{D}$. Denote
the two columns of $\mathbb{D}$ as
${\bf D} = (D_1 , D_2)^{\rm T}$ and $\tilde{\bf D} = \mathbb{E} \,
{\bf D} = (D_2 , -D_1)^{\rm T}$.

The second-order analog to (\ref{eq:2-13}) is
\bea \mathbb{D}^\dagger \mathbb{D} = \tilde{\mathbb{D}}^\dagger
\tilde{\mathbb{D}}  = K \, \mathbb{I} \quad , \quad
\tilde{\mathbb{D}}^\dagger \mathbb{D} = K \, \tilde{\mathbb{I}} \;
, \eeal{eq:2-22}
where $\mathbb{I}$ is the identity, $\tilde{\mathbb{I}}=
\mathbb{E}\mathbb{I} = \mathbb{E}$. Only a weaker version of
(\ref{eq:2-22}) is needed:
\bea \mbox{tr}\; \mathbb{D}^\dagger \mathbb{D} = \mbox{tr}\;
\tilde{\mathbb{D}}^\dagger \tilde{\mathbb{D}} = 2K \quad , \quad
\mbox{tr}\; \tilde{\mathbb{D}}^\dagger \mathbb{D} = 0 \; .
\eeal{eq:2-22w}

The first- and second-order operators are related by
\bea \mathbb{D} + i \tilde{\mathbb{D}} = ( {\bf d} + i \tilde{\bf
d}) \otimes ( {\bf d} + i \tilde{\bf d}) \; , \nonumber
\\
\mathbb{D} = {\bf d} \otimes {\bf d} - \tilde{\bf d} \otimes
\tilde{\bf d} \quad , \quad \tilde{\mathbb{D}} = {\bf d} \otimes
\tilde{\bf d} + \tilde{\bf d} \otimes {\bf d} \; ,\eeal{eq:2-23}
where ${\bf a} \otimes {\bf b}$ has matrix elements $a_ib_j$.
Then (\ref{eq:2-11}) becomes
\bea \mathbb{P} &=& \mathbb{D} \,\Phi^a - \tilde{\mathbb{D}}
\,\Phi^b \; . \eeal{eq:2-25}
To prove this entirely using cartesian components, note that in
analogy to (\ref{eq:2-12}) and (\ref{eq:2-14}),
\bea \varphi^a + i \varphi^b &\equiv& (1/2) \, \mbox{tr}\;
[(\mathbb{D}^\dagger - i \tilde{\mathbb{D}}^\dagger ) \mathbb{P}]
= K\, \left( \Phi^a + i \Phi^b \right) \; , \eeal{eq:2-26}
in which (\ref{eq:2-22w}) has been used. All other steps are
obvious. In analogy to (\ref{eq:formal}), $\varphi^{a,b}$ have no
projection on the null space of $K$.

\subsection{Porting to another plane}
\label{subsect:another}

To port the formalism to any plane with a unit normal $\hat{\bf
n}$, remove explicit reference to $\vnabla_\perp$ by adopting the
equivalent definition (referring to $\vnabla$ but not to
$\vnabla_\perp$)
\bea \tilde{\bf d} = -\hat{\bf n} \times \vnabla \quad , \quad
{\bf d} = \hat{\bf n} \times \tilde{\bf d} \; . \eeal{eq:2-31}
All other formulas remain the same.
However, ${\bf V}$, ${\bf d}$ and $\tilde{\bf d}$ should now be
regarded not as 2-vectors defined on a plane, but as 3-vectors
satisfying a tangential condition: $n_iV_i = n_id_i = n_i
\tilde{d}_i = 0$. Likewise, $\mathbb{P}$, $\mathbb{D}$,
$\tilde{\mathbb{D}}$ are $3 \times 3$ tensors satisfying $n_i
P_{ij} = n_i D_{ij} = n_i \tilde{D}_{ij} = 0$.


\section{Formalism for a sphere}
\label{sect:sphere}


\subsection{Mapping the differential operator}

A \textit{small} part of the surface of a unit sphere can be
regarded as a plane with normal $\hat{\bf n} = {\bf r}$. Then
using (\ref{eq:2-31}), we have, in terms of the angular momentum
${\bf L}$ ($\hbar = 1$):
\bea \tilde{\bf d} = {}-i {\bf L} \quad , \quad  {\bf d} =
\hat{\bf n} \times (-i {\bf L} ) \equiv i\tilde{\bf L} \; .
\eeal{eq:3-01}

\subsection{Vector field}

The planar formula (\ref{eq:2-01a}) applies, but with ${\bf d}$
and $\tilde{\bf d}$ replaced as above. The orthogonality relations
are
\bea {\bf d}^\dagger \cdot {\bf d} = \tilde{\bf d}^\dagger \cdot
\tilde{\bf d} &=& {\bf L}^2 \quad , \quad \tilde{\bf d}^\dagger
\cdot {\bf d} = {\bf L}\cdot \tilde{\bf L} = 0 \; . \eeal{eq:3-02}

Only some key steps in transcribing the formalism will be
sketched; the rest is obvious. For example, (\ref{eq:2-01a})
becomes, upon the replacement (\ref{eq:3-01}),
\bea {\bf V} &=& (i\tilde{\bf L}) \, \Phi^a + (i {\bf L}) \,
\Phi^b \; . \eeal{eq:3a-01}
Define
\bea \varphi^a \equiv (i\tilde{\bf L})^\dagger \cdot {\bf V} = L^2
\Phi^a \quad , \quad  \varphi^b \equiv (i{\bf L})^\dagger \cdot
{\bf V} = L^2 \Phi^b \; . \eeal{eq:3a-02}
Inversion of $L^2$ then gives $\Phi^{a,b}$. The inversion depends
on $\varphi^{a,b}$ having zero projection on the null space of
$L^2$, proved in a manner similar to (\ref{eq:formal}).

\subsection{Polarization field}

Equation (\ref{eq:2-25}) applies with the replacement
(\ref{eq:3-01}), so that (absorbing some signs):
\bea \mathbb{D} \approx ({\bf L} {\otimes} {\bf L}{-} \tilde{\bf
L} {\otimes} \tilde{\bf L}) \quad , \quad \tilde{\mathbb{D}}
\approx (\tilde{\bf L} {\otimes} {\bf L} {+} {\bf L} {\otimes}
\tilde{\bf L}) \; . \eeal{eq:3-03}
The symbol $\approx$ indicates that problems arising out of the
ordering of operators are ignored (see below). The orthogonality
relation (\ref{eq:2-22w}) again holds, except that $K$ and its
eigenvalue are now given by
$K \approx L^4$, $N_K \approx [\ell(\ell{+}1)]^2$.
Everything in the planar case can be repeated with the replacement
of $K$.



The results indicated with $\approx$ are not exact: the order of
the operators leads to ambiguities proportional to the
commutators. Schematically $ [L , L ] = i\hbar L$, so corrections
are ``quantum" and lower-order in $\ell$. In practice, many
experiments targeting the cosmological $B$ mode focus on $\ell
\sim 80$ or larger~\cite{bicep2-1}, so classical (i.e., $\ell \gg
1$) formulas are already useful as a start.

To determine the subsidiary terms, from (\ref{eq:3-03}), one can
write $D_{ij} = X_{ij} - Y_{ij}$ and $\tilde{D}_{ij} = Z_{ij}$,
where to leading order $X_{ij} \approx L_iL_j$, $Y_{ij} \approx
\tilde{L}_i\tilde{L}_j$ and $Z_{ij} \approx L_i \tilde{L}_j$, but
these have to be made (a) symmetric and (b) tangential by adding
terms that are lower-order in ${\bf L}$. The straightforward
construction (\ref{sect:low}) gives
\bea 2X_{ij} &=& L_i L_j {+} L_j L_i + in_i \tilde{L}_j {+} in_j \tilde{L}_i \; , \nonumber\\
2Y_{ij} &=& \tilde{L}_i \tilde{L}_j {+} \tilde{L}_j \tilde{L}_i
-in_i \tilde{L}_j {-}i n_j \tilde{L}_i \; ,
\nonumber \\
2Z_{ij} &=& L_i \tilde{L}_j {+} L_j \tilde{L}_i - in_i L_j {-}
in_j L_i \; . \eeal{eq:a-r1}
The operator $K$ defined by (\ref{eq:2-22w}) and its eigenvalue
$N_K$ now take on the exact values
\bea K = L^2 (L^2-2) \quad , \quad N_K = \ell(\ell{+}1)
[\ell(\ell{+}1)-2] \; , \eeal{eq:a-r2}
consistent with the expression $\nabla^2 (\nabla^2{+}2)$ in
eqn.~(9) in Ref.~\cite{bunn}.

The formalism can be verified by constructing any tangential
traceless symmetric tensor $\mathbb{P}$ on a sphere, and then (a)
evaluating $\varphi^{a,b}$ by (\ref{eq:2-26}); (b) inverting $K$
to obtain $\Phi^{a,b}$; and then (c) recovering $\mathbb{P}$ by
(\ref{eq:2-25}).


\section{Global analysis}
\label{sect:globe}

With global data, the standard mode expansion for scalar functions
leads naturally to corresponding mode expansions for vector and
tensor functions. The modal characteristics allow different
contributions to the polarization (e.g., cosmological versus
foreground contributions) to be disentangled.

\subsection{Planar case}

In the case of a plane, the scalar functions $\Phi^{a,b}$ can be
expanded in terms of plane waves $F_{\bf k} \propto \exp \, (i{\bf
k} \cdot {\bf r})$:
\bea \Phi^{a,b}({\bf r}) &=& \sum_{\bf k}  C^{a,b}_{\bf k} F_{\bf
k}({\bf r}) \; . \eeal{eq:4-01}
Then from (\ref{eq:2-01a}) and (\ref{eq:2-25}), a vector field can
be expressed in terms of ${\bf d} F_{\bf k}$ and $\tilde{\bf d}
F_{\bf k}$ (vector planar harmonics), and a polarization field in
terms of $\mathbb{D} F_{\bf k}$ and $\tilde{\mathbb{D}} F_{\bf k}$
(tensor planar harmonics). The straightforward derivation will not
be shown, except to note that, acting on a plane wave,
$D_1 \mapsto - (k_x^2-k_y^2) = - k^2 \cos2\phi$, $D_2 \mapsto
-2k_xk_y = -k^2\sin2\phi$,
where $\phi$ is the direction of ${\bf k}$, so
\bea {\bf P} = \left(%
\begin{array}{c}
Q \\
U\\
\end{array}%
\right) = -\sum_{\bf k} k^2 \left[   C^a_{\bf k} \left(%
\begin{array}{c}
\cos2\phi \\
\sin2\phi \\
\end{array}%
\right)  +    C^b_{\bf k}
\left(%
\begin{array}{c}
-\sin2\phi \\
\cos2\phi \\
\end{array}%
\right)
 \right] \, F_{\bf k} \; , \phantom{XX} \eeal{eq:4-02}
equivalent to eqn.~(23) in Ref.~\cite{bunn}. Our approach is more
accessible because the planar case is obtained on a stand-alone
basis, and not as a limit of the spherical case.
The polarization direction $\chi$ is given by $\tan 2\chi = U/Q$,
and its relationship with $\phi$ follows trivially from
(\ref{eq:4-02}).
%

\subsection{Spherical case}

On a sphere, the scalar functions can be expanded in spherical
harmonics:
\bea \Phi^{a,b} &=& \sum_{\ell m} C^{a,b}_{\ell m} \, \YLM \; .
\eeal{eq:4-03}
Then a vector field can be expressed in vector spherical harmonics
${\bf Y}^{a,b}_{\ell m}$, and a polarization field in tensor
spherical harmonics $\mathbb{Y}^{a,b}_{\ell m}$:
\bea \left(%
\begin{array}{c}
  {\bf Y}^a_{\ell m}\\
  {\bf Y}^b_{\ell m}\\
\end{array}%
\right) &=&  \frac{1}{\sqrt{\ell(\ell{+}1)} } \,
\left(%
\begin{array}{c}
  i\tilde{\bf L}  \\
  -i{\bf L}  \\
\end{array}
\right) \YLM  \; , \label{eq:VTLM}\\
 \left(%
\begin{array}{c}
  \mathbb{Y}^a_{\ell m}\\
  \mathbb{Y}^b_{\ell m}\\
\end{array}%
\right) &=& \frac{1}{\sqrt{2(\ell^2{+}\ell)(\ell^2{+}\ell{-}2)}}
\left(%
\begin{array}{c}
  \mathbb{D} \\
  \tilde{\mathbb{D}} \\
\end{array}%
\right) \YLM \; , \label{eq:TYLM}\eea
where the prefactor in (\ref{eq:TYLM}) is just $(2N_K)^{-1/2}$
with $N_K$ given by (\ref{eq:a-r2}). The vector spherical
harmonics agree with well-known results~\cite{jack3}. The tensor
harmonics start with $\ell = 2$ because, from (\ref{eq:a-r2}),
$\mathbb{D}$ and $\tilde{\mathbb{D}}$ annihilate $\ell = 0, 1$.


\section{Discussion}
\label{sect:disc}

The representation of linear polarization fields in terms of two
scalar functions involves \textit{second-order} derivatives and
cannot be likened to gradient and curl in the conventional sense.
Nevertheless, a simple analogy exists formally, as shown in the
simplest case by (\ref{eq:2-01}) and (\ref{eq:2-11}). The
transcription of the formalism from a plane to the surface of a
sphere is straightforward.

Vector and tensor fields on a sphere can be more neatly described
using intrinsic coordinates ($\theta$ and $\phi$). Nevertheless
most physicists are more familiar with the cartesian
representation (e.g., $L_x = yp_z - zp_y$). The present account
has kept to the same spirit: embedding in 3-D space but enforcing
tangential conditions. The vector and tensor spherical harmonics
emerge naturally, without invoking the full apparatus of
Newman--Penrose spin-$s$ spherical harmonics~\cite{newman,
campbell}.

In closing, it is useful to recall the steps in the data analysis.
(a) The measured polarization $\mathbb{P}$ on a sphere is
separated into two components, in some sense a generalization of
Helmholtz's theorem: from a spin-1 field to a spin-2 field, and
from a plane to a sphere --- the main focus of the present
pedagogical exposition. (b) With global data, each polarization
component can be decomposed into modes labelled by $(\ell , m)$,
as sketched in Section~\ref{sect:globe}. (c) Finally model
calculations (beyond the purpose of the present paper) would
provide the different characteristics (in mode space, ``map space"
and frequency dependence) of the various dynamical contributions
and thereby allow any \textit{cosmological} contribution to the
$B$ mode to be isolated
--- which is the target of many experiments in progress.

\ack


We thank CS Chu, PT Leung, WT Ni and WM Suen for discussions. The
work of EY is supported by the Kavli Foundation.

\appendix

\section{Lower-order corrections}
\label{sect:low}


The dual is defined as
$\tilde{\bf L} = - \hat{\bf n} \times {\bf L}$,
%
with $\hat{\bf n}$ on the left, to guarantee that $\tilde{\bf L}
\Phi$ are tangential: $n_i \tilde{L}_i \Phi = 0$.

To leading order $X_{ij} \approx L_i L_j$, $Y_{ij} \approx
\tilde{L}_i \tilde{L}_j$.
These expressions when symmetrized
\bea 2X_{ij} \approx L_i L_j + L_j L_i \quad , \quad 2Y_{ij}
\approx \tilde{L}_i \tilde{L}_j + \tilde{L}_j \tilde{L}_i
\eeal{eq:a-03}
are not tangential. To fix this problem, we add correction terms
that are (a) symmetric and (b) lower-order in ${\bf L}$. The only
possibility is a term proportional to
\bea n_i \tilde{L}_j + n_j \tilde{L}_i \; . \eeal{eq:a-04}
All terms in (\ref{eq:a-03}) contain an even number of cross
products (when expressed in terms of position and momenta), so in
(\ref{eq:a-04}) we have to use $\tilde{L}$ and not $L$.

The conditions $n_iX_{ij} = n_i Y_{ij} = 0$ determine the multiple
of (\ref{eq:a-04}) to be added, giving the \textit{exact}
expressions in (\ref{eq:a-r1}). A factor $i$ appears in the
subsidiary terms because ${\bf L}$ and $\tilde{\bf L}$ are
imaginary. Any zero-order terms $\propto n_in_j$ would not be
tangential.

Similarly,
\bea 2Z_{ij} \approx&  L_i \tilde{L}_j + L_j \tilde{L}_i \; ,
\nonumber \eea
%
in which equality is maintained to the accuracy implied by
$\approx$. Enforcing the tangential condition gives the result in
(\ref{eq:a-r1}).
In this case, all terms have an odd number of cross products, so
the subsidiary terms involve $L$ but not $\tilde{L}$.



The three operators $\mathbb{X}$, $\mathbb{Y}$ and $\mathbb{Z}$
are by construction symmetric and tangential, and these properties
are inherited by $\mathbb{D}$ and $\tilde{\mathbb{D}}$. Moreover,
$\mbox{tr} \, \mathbb{X} = \mbox{tr} \, \mathbb{Y} = L^2$,
$\mbox{tr}\, \mathbb{Z} = 0$. Thus $\mathbb{D}$ and
$\tilde{\mathbb{D}}$ are traceless.
Further,
$-\ep_{ijk}n_j (X_{km}-Y_{km}) = Z_{im}$;
i.e., $\tilde{\mathbb{D}} = \mathbb{E}\mathbb{D}$ as expected.



\section*{References}

\begin{thebibliography}{99}


\bibitem{penzias-1}
A.~A.~Penzias and R.~W.~Wilson. A measurement of excess antenna
temperature at 4080 Mc/s. ApJ {\bf 142}, 419--421 (1965). doi:
10.1086/148307

\bibitem{penzias-2}
A.~A.~Penzias and R.~W.~Wilson. Measurement of the flux density of
Cas A at 4080 Mc/s. ApJ {\bf 142}, 1149--1155 (1965). doi:
10.1086/148384

\bibitem{dicke}
R.~H.~Dicke, P.~J.~E.~Peebles, P.~G.~Roll, and D.~T.~Wilkinson.
Cosmic black-body radiation. ApJ {\bf 142}, 414--419 (1965). doi:
10.1086/148306

\bibitem{fixsen-1}
D.~J.~Fixsen and J.~C.~Mather. The spectral results of the
far-infrared absolute spectrophotometer instrument on COBE. ApJ
{\bf 581}, 817--822 (2002). doi: 10.1086/344402

\bibitem{fixsen-2}
D.~J.~Fixsen. The temperature of the cosmic microwave background.
ApJ {\bf 707}, 916 (2009). doi: 10.1088/0004-637X/707/2/916

\bibitem{cobe-first}
G.~F.~Smoot et al. First results of the COBE satellite measurement
of the cosmic microwave background radiation. Adv.\ Space Res.\
{\bf 11} (2), 193--205 (1991).

\bibitem{guth}
A.~H.~Guth. Inflationary universe: A possible solution to the
horizon and flatness problems. Phys.\ Rev.\ D {\bf 23}, 347--356
(1981). doi: 10.1103/PhysRevD.23.347

\bibitem{linde}
A.~D.~Linde. A new inflationary universe scenario: A possible
solution of the horizon, flatness, homogeneity, isotropy and
primordial monopole problems. Physics Letters B, {\bf 108},
389--393 (1982). doi: 10.1016/0370-2693(82)91219-9

\bibitem{wmap}
WMAP Highlights. \verb!https://map.gsfc.nasa.gov!

\bibitem{kogut}
A.~Kogut. WMAP polarization results. New Astronomy Reviews, {\bf
47}, 977--986 (2003). doi: 10.1016/j.newar.2003.09.029

\bibitem{seljak}
U.~Seljak and M.~Zaldarriaga. Signature of gravity waves in the
polarization of microwave background. Phys.\ Rev.\ Lett.\ {\bf
78}, 2054--2057 (1997). doi: 10.1103/PhysRevLett.78.2054

\bibitem{kamkos}
M.~Kamionkowski, A.~Kosowsky and A.~ Stebbins. A probe of
primordial gravity waves and vorticity. Phys.\ Rev.\ Lett.\ {\bf
78}, 2058 (1997). doi: 10.1103/PhysRevLett.78.2058

\bibitem{bicep2-1}
P.~A.~R.~Ade et al. Detection of $B$-mode polarization at degree
angular scales by BICEP2. Phys.\ Rev.\ Lett. {\bf 112}, 241101
(2014). doi: 10.1103/PhysRevLett.112.241101

\bibitem{bicep-planck}
BICEP2/Keck, Planck Collaboration: P.~A.~R.\ Ade et al. A Joint
analysis of BICEP2/Keck array and Planck data. Phys.\ Rev.\ Lett.\
{\bf 114}, 101301 (2015). doi: 10.1103/PhysRevLett.114.101301

\bibitem{planck-foreground}
Planck Collaboration: Y. Akrami et al. Planck intermediate
results. LIV. Polarized dust foregrounds. arXiv:1801.04945v1

\bibitem{progress}
For example, CMB-S4, Simons Observatory, BICEP Array, CHIME,
Litebird, SPIDER2.

\bibitem{cowan} R.~Cowen, Telescope captures view of gravitational waves: Images
of the infant Universe reveal evidence for rapid inflation after
the Big Bang. Nature {\bf 507}, Issue 7492 (2014).\\
\verb!https://www.nature.com/news/telescope-captures-view-of-gravitational-waves-1.14876!

\bibitem{kam1}
M.~Kamionkowski and E.~D.~Kovetz. The quest for $B$ modes from
inflationary gravitational waves. Annual Review of Astronomy and
Astrophysics, {\bf 54}, 227--269 (2016). doi:
10.1146/annurev-astro-081915-023433

\bibitem{bunn}
E.~F.~Bunn, M.~Zaldarriagga, M.~Tegmark and A.~de Oliveira-Costa.
$E/B$ decomposition of finite pixelized CMB maps. Phys.\ Rev.\ D
{\bf 67}, 023501 (2003). doi: 10.1103/PhysRevD.67.023501

\bibitem{jack2}
J.~D.~Jackson, \textit{Classical Electrodynamics}, Second Edition.
Wiley, New York. (1975). Secction 7.2.

\bibitem{soma}
S.~De and H.~Tashiro. Circular polarization of the CMB: a probe of
the first stars. Phys.\ Rev.\ D {\bf 92}, 123506 (2015). doi:
10.1103/PhysRevD.92.123506

\bibitem{king}
S.~King and P.~Lubin. Circular polarization of the CMB:
foregrounds and detection prospects. Phys.\ Rev.\ D {\bf 94},
023501 (2016). doi: 10.1103/PhysRevD.94.023501

\bibitem{chuss}
D.~T.~Chuss et al. Properties of a variable-delay polarization
modulator. Applied Optics {\bf  51} (2), 197--208 (2012). doi:
10.1364/AO.51.000197

\bibitem{class}
L.~Padilla et al. Two-year cosmology large angular scale surveyor
(CLASS) observations: A measurement of circular polarization at 40
GHz. arXiv:1911.00391v1.

\bibitem{bicep1}
H.~C.~Chiang et al. Measurement of cosmic microwave background
polarization power spectra from two years of BICEP data. ApJ {\bf
711}, 1123--1140 (2010). doi: 10.1088/0004-637X/711/2/1123

\bibitem{jack3}
J.~D.~Jackson, \textit{Classical Electrodynamics}, Second Edition.
Wiley, New York. (1975). Section 16.2.


\bibitem{newman}
E.~T.~Newman and R.~Penrose. Ten exact gravitationally-conserved
quantities. Phys.\ Rev/.\ Lett.\ {\bf 15}, 231--233 (1965). doi:
10.1103/PhysRevLett.15.231

\bibitem{campbell}
W.~B.~Campbell. Tensor and spinor spherical harmonics and the
spin-$s$ harmonics ${}_sY_{\ell m}(\theta,\phi)$. J.\ Math.\
Phys.\ {\bf 12}, 1763--1770 (1971). doi: 10.1063/1.1665802


\end{thebibliography}
\end{document}